\documentclass[aps,pra,twocolumn,superscriptaddress,floatfix]{revtex4}
\usepackage{amsmath,amssymb,graphicx,graphics,color}
\usepackage{dcolumn}
\usepackage{bm}
\usepackage{psfrag}
\usepackage{epstopdf}

\begin{document}

\title{Spin-orbit coupled fermions in ladder-like optical lattices at half-filling}
\author{G. Sun}
\affiliation{Institut f\"ur Theoretische Physik, Leibniz Universit\"at Hannover, 30167~Hannover, Germany}

\author {J. Jaramillo}
\affiliation{Institut f\"ur Theoretische Physik, Leibniz Universit\"at Hannover, 30167~Hannover, Germany}

\author{L. Santos}
\affiliation{Institut f\"ur Theoretische Physik, Leibniz Universit\"at Hannover, 30167~Hannover, Germany}
 
\author {T. Vekua}
\affiliation{Institut f\"ur Theoretische Physik, Leibniz Universit\"at Hannover, 30167~Hannover, Germany}

\begin{abstract}
We study the ground-state phase diagram of two-component fermions loaded in a ladder-like lattice at half filling 
in the presence of spin-orbit coupling. 
For repulsive fermions with unidirectional spin-orbit coupling along the legs we identify a N\'{e}el state which is separated from rung-singlet 
and ferromagnetic states by Ising phase transition lines. These lines cross for maximal spin-orbit coupling and a direct Gaussian phase 
transition between rung-singlet and ferro phases is realized. For the case of Rashba-like spin-orbit coupling, besides the rung singlet phases two distinct striped ferromagnetic phases are formed. In case of attractive fermions with spin-orbit coupling at half-filling 
 for decoupled chains we identify a dimerized state that separates a singlet superconductor and a ferromagnetic states. 
\end{abstract}

\maketitle

\date{\today}




\section{Introduction}

The possibility of inducing synthetic electromagnetism in ultra-cold gases has attracted recently a large deal of attention. 
In spite of the electric charge neutrality of an atom, synthetic magnetic field may be induced by a proper laser arrangement~\cite{Spielman1}. 
Interestingly, uni-directional spin-orbit coupling~(USOC) resulting from an equal superposition of Rashba~\cite{Rashba} and linear Dresselhaus~\cite{Dresselhaus} terms, 
has been realized for both spinor Bose~\cite{Spielman} and Fermi gases~\cite{spinorbitfermions, Cheuk} with the help of counter-propagating Raman lasers. 
Recently this technique has allowed for the observation of superfluid Hall effect~\cite{LeBlanc2012}, Zitterbewegung~\cite{Zitterbewegung}, and
the spin-Hall effect in a quantum gas~\cite{Beeler2013}. 
Several theory works have discussed the creation of pure Rashba or Dresselhaus SOC with optical~\cite{Campbell2011} and magnetic means~\cite{Anderson2013}, 
and even proposed methods to generate a three-dimensional SOC~\cite{Anderson2012}.

The presence of a synthetic SOC is expected to lead to a rich physics for atoms loaded in optical lattices. 
For two-dimensional Hubbard models at half filling the effects of a Rashba-like SOC were studied both for two-component bosons and fermions, for which
exotic spin textures in the ground state such as coplanar spiral waves and stripes as well as non-coplanar vortex/antivortex configurations have been predicted~\cite{Cai, Cole, Radic, Gong}. 
Note, however, that the SOC introduces frustration, invalidating quantum Monte Carlo~(MC) approaches, and hence most studies have relied on classical MC calculations.

In this paper we analyze the effects of SOC in a two-component Fermi gas loaded in an optical lattice in the Mott-insulator regime. 
Since we are interested in the quantum spin-$1/2$ phases in the presence of SOC, we can not rely on classical MC, and must hence employ 
exact diagonalization or density-matrix renormalization group~(DMRG) techniques. We employ the latter in our paper, restricting 
our analysis to the minimal system where the non-Abelian 
character of the vector potential may be manifested allowing non-trivial effects of SOC without the need of breaking the time-reversal invariance, namely a two-leg ladder-like optical lattice, which may be created by incoherently combining a 1D lattice and a two-well potential.
By a combination of numerical DMRG results, bosonization techniques and strong rung-coupling expansions, we obtain the 
spin quantum phases for both a USOC with different orientations with respect to the ladder, and the isotropic SOC.
 
The paper is organized as follows. In Sec.~\ref{sec:Model} we introduce the effective spin model for a Mott state of two-component fermions with USOC in a ladder-like lattice. 
In Sec.~\ref{sec:Decoupled} we review the phases for the case of decoupled one-dimensional lattices. Section~\ref{sec:TwoLeg} deals with the quantum phases of an USOC 
discussing the different orientations between the USOC and the ladder legs. In Sec.~\ref{sec:NonAbelian} we analyze the case of an isotropic SOC. We finally summarize in Sec.~\ref{sec:Conclusions}.

\begin{figure}[t]
\includegraphics[width=0.8\columnwidth]{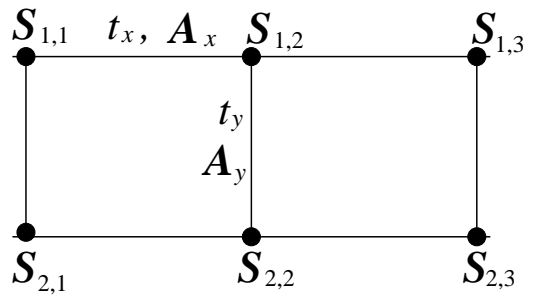}
\caption{Two-leg ladder lattice of $s=1/2$ spins ${\mathbf S}_{\alpha,j}$, where $\alpha=1,2$ enumerates the ladder legs, and $j=1,2,3,\cdots, L$ labels the ladder rungs.}
\vspace*{-0.3cm}
\label{fig:SpinLadder}
\end{figure}

\section{Effective spin model for two-component fermions with USOC}
\label{sec:Model}

Recent experiments have realized an USOC characterized by a Hamiltonian of the form~\cite{Spielman}:
\begin{equation}
\label{synthetic}
 H_{USOC}=\frac{1}{2m}({\mathbf p}\sigma^0-{\mathbf A})^2+\frac{\delta }{2}\sigma^z-\frac{h}{2}\sigma^x,
\end{equation}
where $\sigma^{z,x}$ are Pauli matrices, $\sigma^0$ is the identity matrix, and the effective vector potential 
for counter-propagating Raman lasers on the $xy$ plane is given by 
${\mathbf A}=-\hbar{\bf k}_0\sigma^z$, with ${\bf k}_0=(k_0^x,k_0^y,0)$.
Here the eigenvectors of $\sigma^z$ correspond to atomic hyperfine components, the 
term $\frac{\delta}{2}\sigma^z$ is due to detuning from resonance, and $h$ is the Rabi coupling.
Crucially, ${\mathbf A}$ cannot be completely gauged out, since it 
does not commute with the scalar potential $\Phi=\frac{\delta}{2}\sigma^z-\frac{h}{2}\sigma^x$.

We consider a two-component Fermi gas loaded in a ladder-like optical lattice of inter-site spacing $a$, with the ladder legs oriented along $x$ and the rungs along $y$.
Projecting on the lowest lattice band~\cite{Jaksch} one obtains, in absence of SOC, the two-component Fermi-Hubbard model:
\begin{equation}
\label{Mott}
H_{FH}=-\!\!\!\!\!\sum_{(i,i'),\sigma,\sigma'}\!\!\! t_{i,i'}\sigma^0_{\sigma,\sigma'}a^{\dagger}_{i,\sigma}a_{i',\sigma'}+\frac{U}{2}\sum_i n_i(n_i-1),
\end{equation}
where $a_{i,\sigma}$ is the annihilation operator of fermions with spin $\sigma= \uparrow,\downarrow$ on site $i$,  
$n_i=\sum_{\sigma}a^{\dagger}_{i,\sigma}a_{i,\sigma}$, $U$ characterizes the on-site interaction, and 
$t_{i,i'}$ are the hopping amplitudes along the bonds connecting nearest-neighbor sites $(i,i')$, with the hopping along the legs~(the rungs) given by $t_{i,i'}=t_x$~($t_y$).
The presence of SOC results in the Peierls substitution
 $t_{i,i'}\sigma^0 \to t_{i,i'} e^{i\frac{ {\mathbf A}({\mathbf r}_{i'}-{\mathbf r}_i)}{\hbar}}$. 
In the strong coupling limit, $U\to \infty$, and considering half-filling (i.e. we consider a Mott phase with one fermion per site~\cite{Jordens2008,Schneider2008}), 
the Fermi-Hubbard model may be re-written as an effective spin-$1/2$ model of the form:
\begin{eqnarray}
\label{2D}
&&\!H\!\!=\!\! J_{\parallel}\sum_{\alpha,j} \! \Big \{ \cos(2k_0^x a) \mathbf{S}_{\alpha,j}  \mathbf{S}_{\alpha,j+1}
 \nonumber \\
&&\! +2\sin^2(k^x_0 a) S^z_{\alpha,j} S^z_{\alpha,j+1}+ \sin (2k^x_0 a) [\mathbf{S}_{\alpha,j } \times \mathbf{S}_{\alpha,j+1} ]^z \Big \} \nonumber\\
&&\!\! +J_{\bot}\sum_{j} \Big\{ \cos(2k^y_0 a) \mathbf{S}_{1,j}  \mathbf{S}_{2,j}
 \nonumber\\  
&&+2\sin^2(k^y_0 a) S^z_{1,j} S^z_{2,j} +\!  \sin {2k^y_0 a}[\mathbf{S}_{1,j} \times \mathbf{S}_{2,j} ]^z \Big\} \nonumber \\
&&+\delta \sum_{\alpha,j}  S^z_{\alpha,j}-h  \sum_{\alpha,j}  S^x_{\alpha,j}  . 
\end{eqnarray}

 where $J_{\parallel}= {4t_x^2}/{U}$, $J_{\bot}= {4t_y^2}/{U}$,  and
  \begin{equation}
\mathbf {S}_{\alpha,j}= (a^{\dagger}_{ \alpha,j,\uparrow},a^{\dagger}_{\alpha,j  ,\downarrow})   \frac{\boldsymbol {\sigma}}{2}  \begin{pmatrix} a_{\alpha,j  ,\uparrow}\\ a_{ \alpha,j,\downarrow} \end{pmatrix}.
\end{equation}
are the spin operators associated to the leg $\alpha=1,2$ and the rung $j$~(see Fig.~\ref{fig:SpinLadder}), with the site index $i$ in Eq.~(\ref{Mott}) split into leg and rung indices: $i\to (\alpha, j)$.
The value of $k_0^x$ and $k_0^y$ is provided by the orientation between the Raman lasers creating the USOC and the ladder axis.
Note that the scalar potential $\Phi$ produces the last two terms in Eq.~(\ref{2D}), whereas the 
vector potential ${\bf A}$ produces Dzyaloshinskii-Moriya~(DM) terms~\cite{Dzyaloshinskii,Moriya}, 
$\sim  [\mathbf{S}_{\alpha,j } \times \mathbf{S}_{\alpha',j'} ]^z$, 
as well as easy-axis anisotropy~(EAA) along ${\bf e}_z$.

\section{Decoupled chains}
\label{sec:Decoupled}

\subsection{Repulsive interactions}

\begin{figure}[t]
\includegraphics[width=0.8\columnwidth]{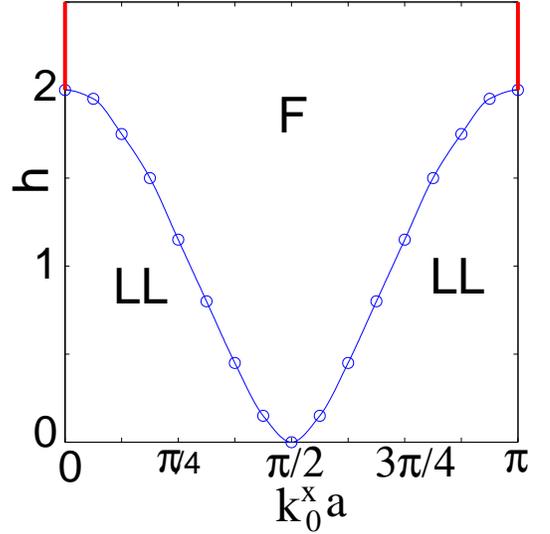}\\

\caption{ (Color online) Ground states of  a 1D spin-1/2 chain with USOC and transverse magnetic field obtained using DMRG for $96$ sites. The magnetic field is in units of $J_{\parallel}$. LL denotes a luttinger liquid phase and F stands for ferromagnetic state. 
}
\vspace*{-0.3cm}
\label{fig:repulsivechain}
\end{figure}

We first discuss the case of decoupled chains, $J_{\bot}=0$~(i.e. $t_y=0$), which results in the 1D Hamiltonian
\begin{eqnarray}
\label{chain}
H_{1D}&=& J_{\parallel}\sum_j \Big( S^z_j S^z_{j+1}+\cos{2k^x_0 a}( S^x_j S^x_{j+1}+S^y_j S^y_{j+1} )\nonumber\\
&{}&+\sin {2k^x_0 a}( S^x_j S^y_{j+1}-S^y_j S^x_{j+1} )\Big)- h \sum_j  S_j^x .
\end{eqnarray}
For $k_x^0=0$  Eq.~(\ref{chain}) describes an SU($2$)-symmetric spin-$1/2$ antiferromagnetic chain in external magnetic field, which 
is exactly solvable by means of Bethe ansatz~\cite{Takahashi}. The ground state is a gapless Luttinger liquid~(LL) 
for $h< 2J_{\parallel}$, and a fully polarized state for  $h> 2J_{\parallel}$. 
These two phases are separated by a commensurate-incommensurate (C-IC) phase transition. 

In order to discuss the effects of the USOC it is convenient to introduce a gauge transformation 
that renders exchange interactions explicitly SU($2$) invariant, 
$H_{1D}\to UH_{1D}U^{\dagger}= \bar H_{1D}$, 
 where 
$U=\prod_j  e^{-2ik^x_0a jS_j^z}$. The spin operators transform as
\begin{eqnarray}
\label{gaugeinspins}
\bar S^x_j &=&  \cos{(2k^x_0 a j)} S_j^x -\sin{(2k^x_0 a j)} S^y_j ,\nonumber\\
 \bar S^y_j&=&  \cos{(2k^x_0 a j)} S_j^y +\sin{(2k^x_0 a j) }S^x_j .
\end{eqnarray}
and $\bar S^z_j= S^z_j$, and the Hamiltonian becomes
\begin{equation}
\label{effectiverotated}
\bar H_{1D}= J_{\parallel}\!\sum_j\mathbf {\bar S}_j\mathbf {\bar S}_{j+1}-\sum_j {\bf h}_j({\bf k}_0) \mathbf {\bar S}_j,
\end{equation}
where the effect of the USOC is entirely absorbed into an external magnetic field, 
${\bf h}_j({\bf k}_0)=h(\cos(2k^x_0 a j),\sin(2k^x_0 a j),0)$, that spirals on the $xy$ plane.

For $k^x_0 a=\pi/2$, ${\mathbf h}_j=(-1)^j \, h\, {\bf e}_x$, i.e. a staggered effective magnetic field. 
A staggered field constitutes a relevant perturbation~(in the renormalization group sense) as it couples to the 
N\'{e}el order, which is one of the leading instabilities in a 1D antiferromagnetic chain. As a result of that,  
a gap in the excitation spectrum, $\Delta E\sim h^{2/3}$, opens for any arbitrary coupling $h$. 
The low-energy behavior is described by a massive sine-Gordon model 
where one of the breather modes is degenerate with soliton and anti-soliton excitations~\cite{Oshikawa}.
In the gauge transformed variables the ground state developes N\'{e}el order, which after un-doing the 
gauge transformation results for the original spin operators into an uniformly magnetized state, i.e. a 
ferromagnetic~(F) state, although magnetization is never fully saturated for $k_0^x\neq 0$.

\begin{figure}[t]
\includegraphics[width=0.8\columnwidth]{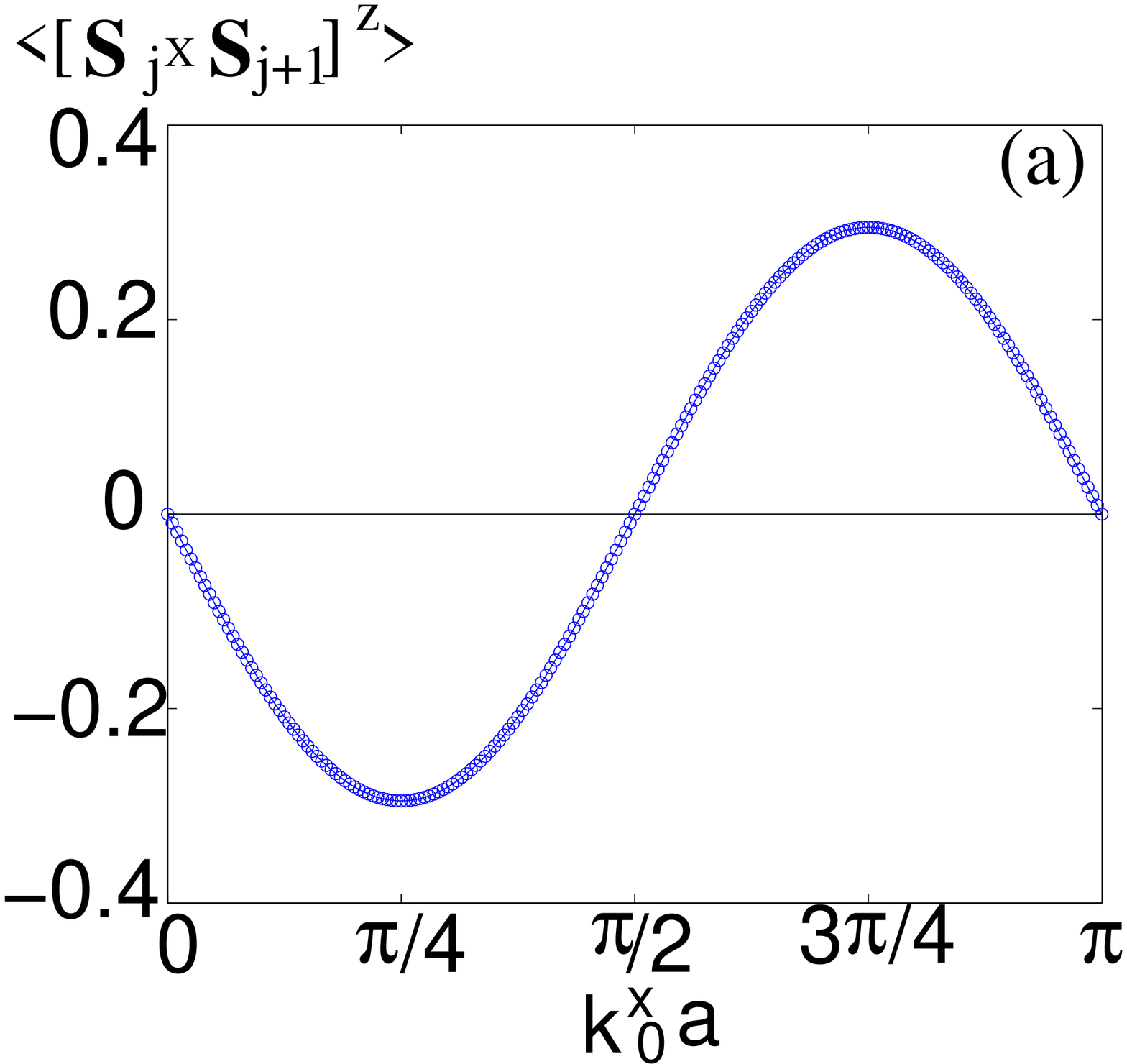}
\includegraphics[width=0.8\columnwidth]{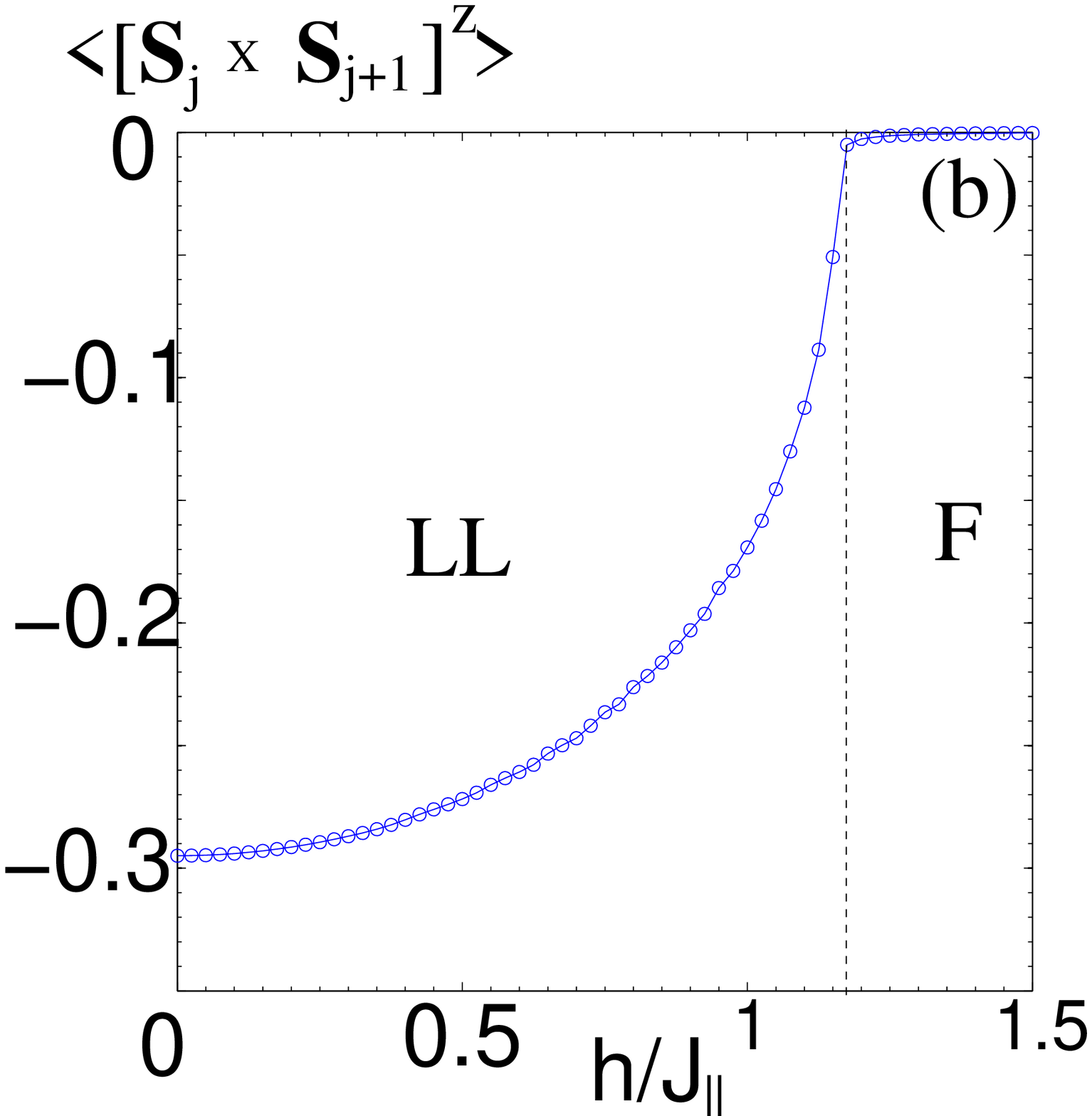}
\caption{ Expectation value of the vector product of two neighboring spins as a function of: (a) the USOC parameter, $k_0^xa$, for $h=0$; and (b) 
magnetic field for $k_0^x a=\pi/4$.}
\label{fig:chirality}
\end{figure}

For  $0<k^x_0 a<\pi/2$, ${\bf h}_j({\bf k}_0)$ is incommensurate and hence 
the gapless LL phase survives up to a finite $h$ value at which the F phase is reached.
We have employed the matrix product formulation~\cite{Verstraete} of DMRG method \cite{White, Uli} to obtain numerically 
the phase diagram for arbitrary values of the USOC~(see Fig.~\ref{fig:repulsivechain}). 
This phase diagram confirms the existence of a gapless LL and a gapped F phase separated by a C-IC transition. 
Note that correlation functions, which decay algebraically in the LL phase and exponentially in the F phase, 
are generically incommensurate due to the DM anisotropy and the vector product of two neighbouring spins has finite expectation value
$\langle [\mathbf{S}_{j} \times \mathbf{S}_{j+1} ]^z  \rangle \sim -\sin (2k^x_0 a) $ as depicted in Fig.~\ref{fig:chirality}(a). Its magnetic field dependence is presented in Fig.~\ref{fig:chirality}(b).

\subsection{Attractive interactions}
\begin{figure}[t]
\vspace*{0.4cm}
\includegraphics[width=0.8\columnwidth]{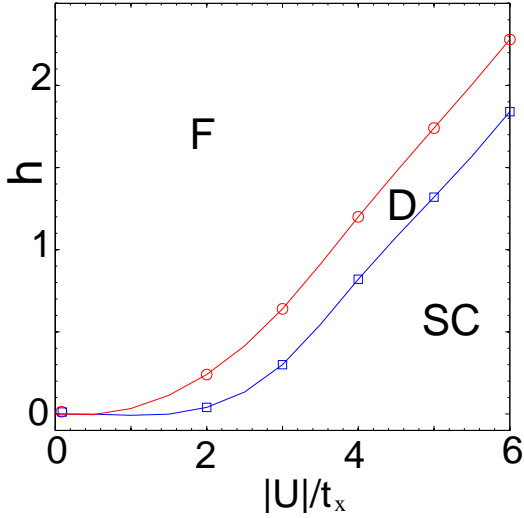}
\caption{ Phase diagram, for an attractive two-component Fermi Hubbard model on a chain at half filling with maximal USOC, where  D denotes a dimerized phase, and SC stands for a 1D superconductor. The magnetic field is in units of $t_x$. The phase boundaries are obtained after finite-size extrapolation from data obtained for $128$, $256$, $512$ and $1024$ sites.}
\vspace*{-0.3cm}
\label{fig:attractive}
\end{figure}

For the decoupled chains we have also studied the case of two-component fermions with attractive interactions. 
The most interesting ground-state physics occurs at half-filling in the vicinity of the maximal USOC, $k^x_0 a\simeq \pi/2$. 
In this case, after particle-hole transformation the 1D Fermi-Hubbard model becomes dual to the repulsive ionic-Hubbard model~\cite{Fabrizio}, 
being characterized by the existence of a dimerized~(D) phase between a superconducting~(SC) phase and the F state.
With increasing magnetic field the SC phase undergoes a Kosterlitz-Thouless~(KT) transition into the D state, 
where translational symmetry is spontaneously broken. Further increasing the magnetic field results in a D-F Ising transition. 
We characterized the D phase in our numerical simulations by means of the dimerization order parameter, which 
in a chain with $L$ sites is defined as:
\begin{equation}
D=\sum_j  \frac{(-1)^j}{L} \langle  a^{\dagger}_{j,\uparrow} a_{j+1,\downarrow}-  a^{\dagger}_{j,\downarrow} a_{j+1,\uparrow} +h.c. \rangle. 
\end{equation}
The phase diagram of the 1D attractive Fermi-Hubbard model with $k^x_0 a=\pi/2$ at half-filling is presented in Fig.~\ref{fig:attractive}.


\section{Two-leg ladder with USOC}
\label{sec:TwoLeg}

We consider now the case of coupled chains with nonzero hoppings $t_{x,y}$.
As mentioned above, the value of $k_x^0$ and $k_y^0$ depends on the orientation of the USOC lasers and the 
ladder axis. In the following we consider separately the case in which the USOC is along the rungs and 
that in which the USOC is along the legs.

\subsection{USOC along the ladder rungs}
\label{subsec:USOC-rungs}

\begin{figure}[t]
\vspace*{0.3cm}
\includegraphics[angle=90,width=7.0cm]{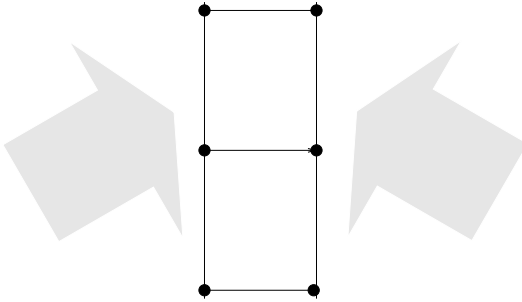}
\caption{  Raman lasers counter-propagating along ladder rungs result in USOC as discussed in subsection~\ref{subsec:USOC-rungs}.}
\vspace*{-0.3cm}
\label{fig:laserRungs}
\end{figure}

We analyze first the case of an USOC along the ladder rungs, i.e. $k^x_0=0$ in Eq.~(\ref{2D}). 
For $k^y_0= 0$ the magnetic field introduces two C-IC phase transitions: from a rung-singlet~(RS) into a LL and then from the 
LL into the fully polarized F state.  
As in our discussion of Sec.~\ref{sec:Decoupled} it is convenient to introduce the gauge transformation
\begin{eqnarray}
\label{gaugeinspinsRungs}
  \bar S^x_{\alpha, j}& =& \cos{(2k^y_0 a \alpha)} S_{\alpha, j}^x -\sin{(2k^y_0 a \alpha)} S^y_{\alpha, j} ,\nonumber\\
 \bar S^y_{\alpha, j}&=&  \cos{(2k^y_0 a \alpha)} S_{\alpha, j}^y +\sin{(2k^y_0 a \alpha) }S^x_{\alpha, j} 
\end{eqnarray}
and $ \bar S^z_{\alpha, j}= S^z_{\alpha,j}$. For the case of the maximal USOC, $k^y_0a=\pi/2$, the 
gauge transformed Hamiltonian becomes:
\begin{eqnarray}
\label{RStransforme}
\bar H&=& J_{\parallel}\!\sum_{\alpha=(1,2),j}\mathbf {\bar S}_{\alpha,j}\mathbf {\bar S}_{\alpha,j+1} +  J_{\bot}\!\sum_j\mathbf {\bar S}_{1,j}\mathbf {\bar S}_{2,j} \nonumber\\  
&-&h \! \!\!\! \sum_{\alpha=(1,2),j}\!(-1)^{\alpha} \bar S_{\alpha,j}^x .
\end{eqnarray}
In the strong rung-coupling limit, $J_{\bot}\gg J_{\parallel}$, the ground state becomes a rung-product state of the form:
\begin{equation}
\label{rung-singlet}
|\bar {RS}\rangle = \prod_j\Big(|\bar \uparrow_{1,j}\rangle \otimes|\bar \downarrow_{2,j}\rangle  -
\beta |\bar \downarrow_{1,j}\rangle \otimes|\bar \uparrow_{2,j}\rangle \Big)/\sqrt{1+\beta^2},
\end{equation}
where $\{ \bar \uparrow, \bar \downarrow\}$ refer to the eigenstates of $\bar S^x$.
For $h=0$, $\beta=1$ and the ground-state is a product state of singlets along the rungs. 
With increasing magnetic field $\beta$ decreases gradually tending to zero. For $\beta=0$ the ground-state after undoing 
the gauge transformation translates into the F state.
Hence, for  $k^y_0a=\pi/2$ the magnetic field just results in an adiabatic evolution of $|\bar {RS}\rangle$ into the F state.

To address the general case $0<k_0^y< \pi/2$ we consider the case of weak USOC, $k_0^ya\ll 1$, 
closely following the strong rung-coupling derivation of Ref.~\cite{Penc}.
For $h=0$ the ground state is well approximated by a direct product of singlets along the rungs, and the 
energy gap to the lowest rung triplet excitation is $\sim J_{\bot}$. 
The external magnetic field splits linearly the rung triplet excitations, 
and the energy of the state where both spins of the rung point in the direction of the field approaches that of the RS state for $h\sim  J_{\bot}$.   
Identifying the RS state on a rung with an effective spin-$1/2$ pointing down, and the $S^x=1$ component of the rung triplet state 
with the spin-$1/2$ pointing up, the effective pseudo-spin-$1/2$ model in the strong rung-coupling limit for $h\sim J_{\bot}$ takes the form 
of an XXZ model in a tilted uniform magnetic field:
\begin{eqnarray}
\label{effs}
H_{\tau}&=&J_{\parallel} \sum_j (\frac{1}{2}\tau^x_j\tau^x_{j+1}+\tau^y_j\tau^y_{j+1}+   \tau^z_j\tau^z_{j+1})\nonumber\\
&-&h_x\sum_j \tau^x_j- h_y\sum_j \tau^y_j 
\end{eqnarray}
where $\tau^{x,y,z}$ are the pseudo-spin-$1/2$ operators, 
$h_x=h-J_{\bot} \cos{2k_0^y a} +J_{\bot}(1- \cos{2k_0^y a})/4   -J_{\parallel}/2$, and $h_y= J_{\bot} \sin{2k_0^y a}/\sqrt{2} $.
With varying $h_x$ the model~\eqref{effs} undergoes changes in three ground-state phases~\cite{Ovchinnikov}: two F phases separated by Ising transitions 
from an intermediate N\'{e}el phase in $\tau^z$ state. One of the F phases of the effective model~(\ref{effs}) translates to the RS phase of the ladder, whereas the N\'{e}el phase and the second F phase of~(\ref{effs}) 
translate into identical ladder phases.  Note that it is the DM interaction that in the leading order breaks in Eq.~(\ref{effs}) 
the $U(1)$ rotation symmetry in the $yz$ plane allowing for the N\'{e}el ordering. 

We consider at this point weakly coupled chains, $J_{\bot}\ll J_{\parallel}$, again for weak USOC, $k_0^ya\ll 1$. 
For this case we can use bosonization mapping~\cite{Gogolin} with the convention:
\begin{eqnarray}
\label{bosonizationRules}
S_{\alpha,j}^x& \to& \frac{\partial_x \phi_{\alpha}}{\sqrt{2\pi}}+(-1)^j \sin {\sqrt{2\pi } \phi_{\alpha}}, \nonumber \\
S_{j,\alpha}^y & \to & (-1)^j\!\sin{  \sqrt{2\pi } \theta_{\alpha} }+\!\cdots  \\
S_{\alpha,j}^z\! &\to& \!\! (-1)^j\!\cos{  \sqrt{2\pi } \theta_{\alpha} }\!+\!\cdots  \nonumber
\end{eqnarray}
where $x=ja$, the dots denote sub-leading fluctuations of uniform components, and we have introduced two pairs of dual bosonic fields,
 $[\theta_{\alpha}(x),\partial_y \phi_{\alpha'}]=i\delta_{\alpha,\alpha'}\delta(x-y)$. 
It is convenient to introduce the symmetric and antisymmetric combinations of the original bosonic fields, 
$\theta_\pm=(\theta_1\pm \theta_2)/\sqrt{2}$, $\phi_\pm=(\phi_1\pm \phi_2)/\sqrt{2}$.
We treat $J_{\bot}$ as a perturbation of the two decoupled chains retaining only the relevant contributions that it generates.  
We obtain the following Hamiltonian density:
\begin{eqnarray}
\label{effectivetwocomponent1}
{\mathcal H_B}&=&\sum_{\nu=\pm}\frac{ v_{\nu}}{2}\left[ (\partial_x \phi_{\nu})^2          + (\partial_x \theta_{\nu})^2  \right] -h\partial_x\phi_+/\sqrt{\pi} \nonumber \\
&+& \tilde J_{\bot}  ( 2 \cos \sqrt{4\pi}\theta_- - \cos\sqrt{4\pi} \phi_{+}  + \cos\sqrt{4\pi} \phi_{-}   )\nonumber \\
&+&\tilde d_{\bot}\left( \cos \sqrt{\pi} \phi_+ \sin{\sqrt{\pi}\theta_+}  \sin \sqrt{\pi} \phi_- \cos{\sqrt{\pi}\theta_-}  \right. \nonumber\\
&-&\left. \sin \sqrt{\pi} \phi_+ \cos{\sqrt{\pi}\theta_+}  \cos \sqrt{\pi} \phi_- \sin{\sqrt{\pi}\theta_-}  \right)\nonumber\\
&+&   d_{\bot} ( \cos \sqrt{4\pi}\theta_- + \cos \sqrt{4\pi}\theta_+ )   ,
\end{eqnarray}
where 
$$
v_{\pm}\simeq \frac{J_{\parallel}a\pi}{2}\left (1\pm \frac{J_{\bot} \cos(2k_0^y a) }{J_{\parallel} \pi^2} \right ) 
$$
and $\tilde J_{\bot}\sim J_{\bot} \cos (2k_0^ya)$.  
The DM anisotropy generates the terms with the coupling constant
$\tilde d_{\bot}\sim J_{\bot} \sin (2k_0^ya)$, whereas the EEA induces the terms with the pre factor $d_{\bot}\sim J_{\bot}(1-\cos(2k_0^ya))$. 
All three factors $\tilde J_{\bot}$,  $\tilde d_{\bot}$ and $d_{\bot}$ depend as well on a short-distance cut-off.
Note that the magnetic field, $h$, just couples to the symmetric sector.

For $k_0^ya\ll 1$, $d_{\bot}, \tilde d_{\bot} \ll \tilde J_{\bot}$, and the 
antisymmetric sector remains gapped with $\langle \theta_-\rangle =\sqrt{\pi}/2$.
Note that when the magnetic field suppresses the dominant coupling in the symmetric sector, $\tilde J_{\bot}\cos{\sqrt{4\pi}\phi_+}$, 
the EEA term induces the leading instability.

\begin{figure}[t]
\vspace*{0.3cm}
\includegraphics[width=5.0cm]{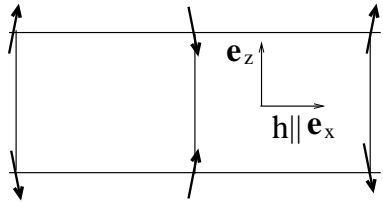}
\caption{ N\'{e}el state configuration for USOC along rungs. 
The inter-leg correlation functions are also antiferromagnetic,  $\langle S^z_{1,i} S^z_{2,j} \rangle\sim (-1)^{i-j+1}$.}
\vspace*{-0.3cm}
\label{fig:Neel}
\end{figure}

The symmetric sector can be solved by means of the Jordan-Wigner mapping and a subsequent Bogoliubov transformation. 
It supports two Ising phase transitions with increasing magnetic field that separate three different ground-state phases.
In the original spin variables, these phases translate into the RS phase, a  N\'{e}el state with 
order parameter $n=(-1)^{j+\alpha}\langle S_{\alpha,j}^z \rangle\neq 0$, and the F phase. 
In N\'{e}el state spins are canted uniformly along the applied field as depicted in Fig.~\ref{fig:Neel}.

\begin{figure}[t]
\vspace*{0.3cm}
\includegraphics[width=0.8\columnwidth]{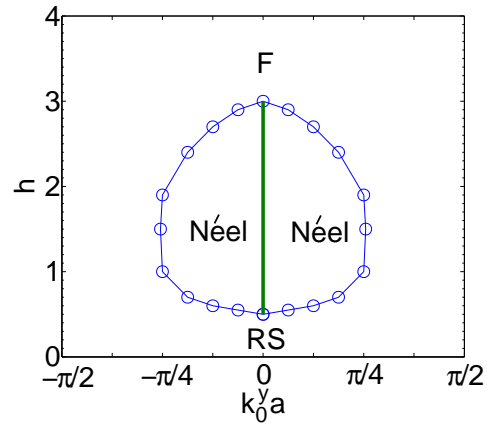}
\caption{ Phase diagram for the USOC along ladder rungs. The phase transition curve into the N\'{e}el state is determined from 
the closing of the gap between the two lowest eigenstates, see Fig.~\ref{fig:squareNeelRung}(b). For $k_0^y=0$ a LL line is realized
between the RS and F phases. The magnetic field is in units of $J_{\parallel}=J_{\bot}$.}
\vspace*{-0.3cm} 
\label{fig:USOCrungs}
\end{figure}

\begin{figure}[t]
\vspace*{0.3cm}
\includegraphics[width=0.8\columnwidth]{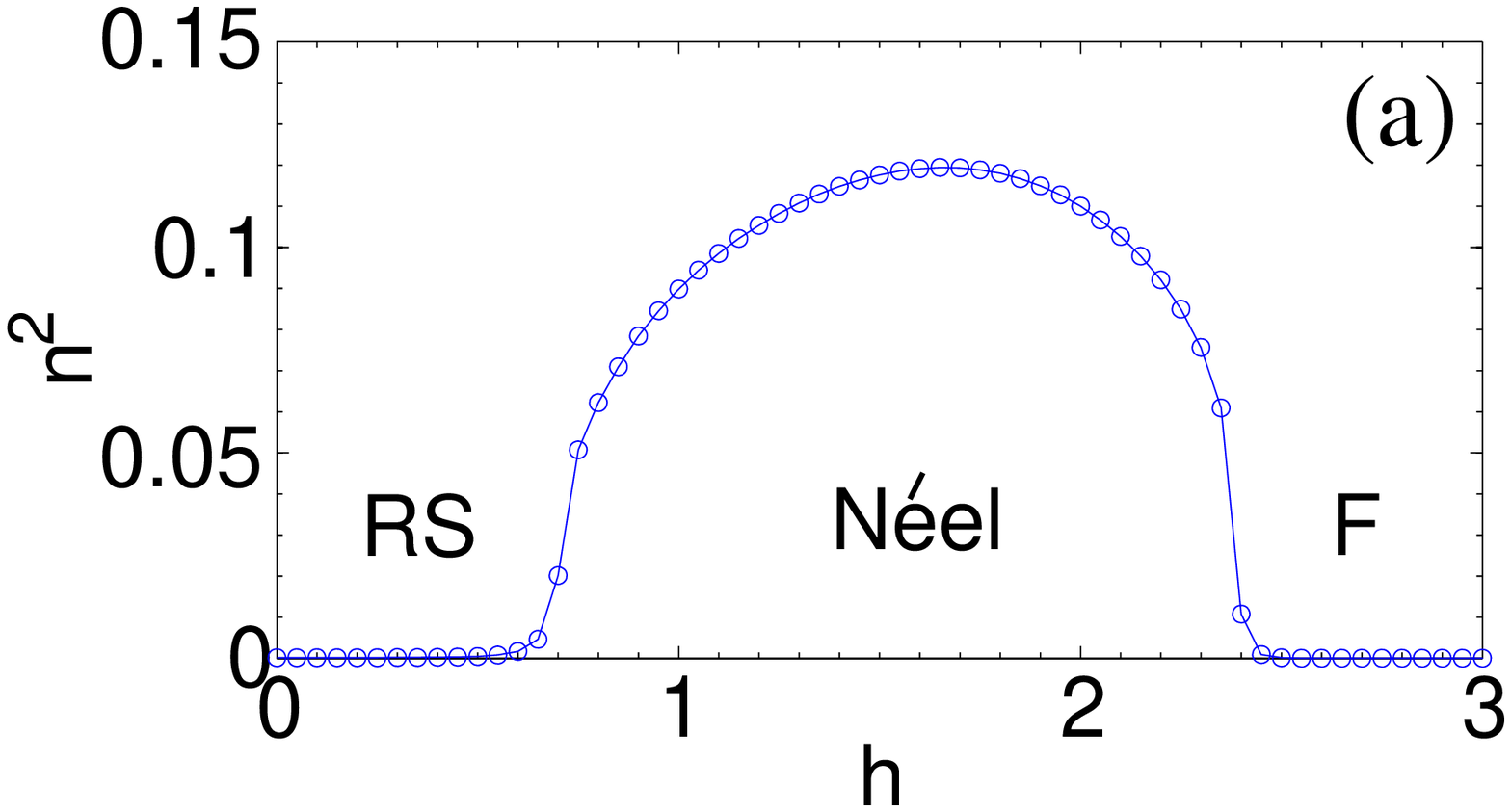}
\includegraphics[width=0.8\columnwidth]{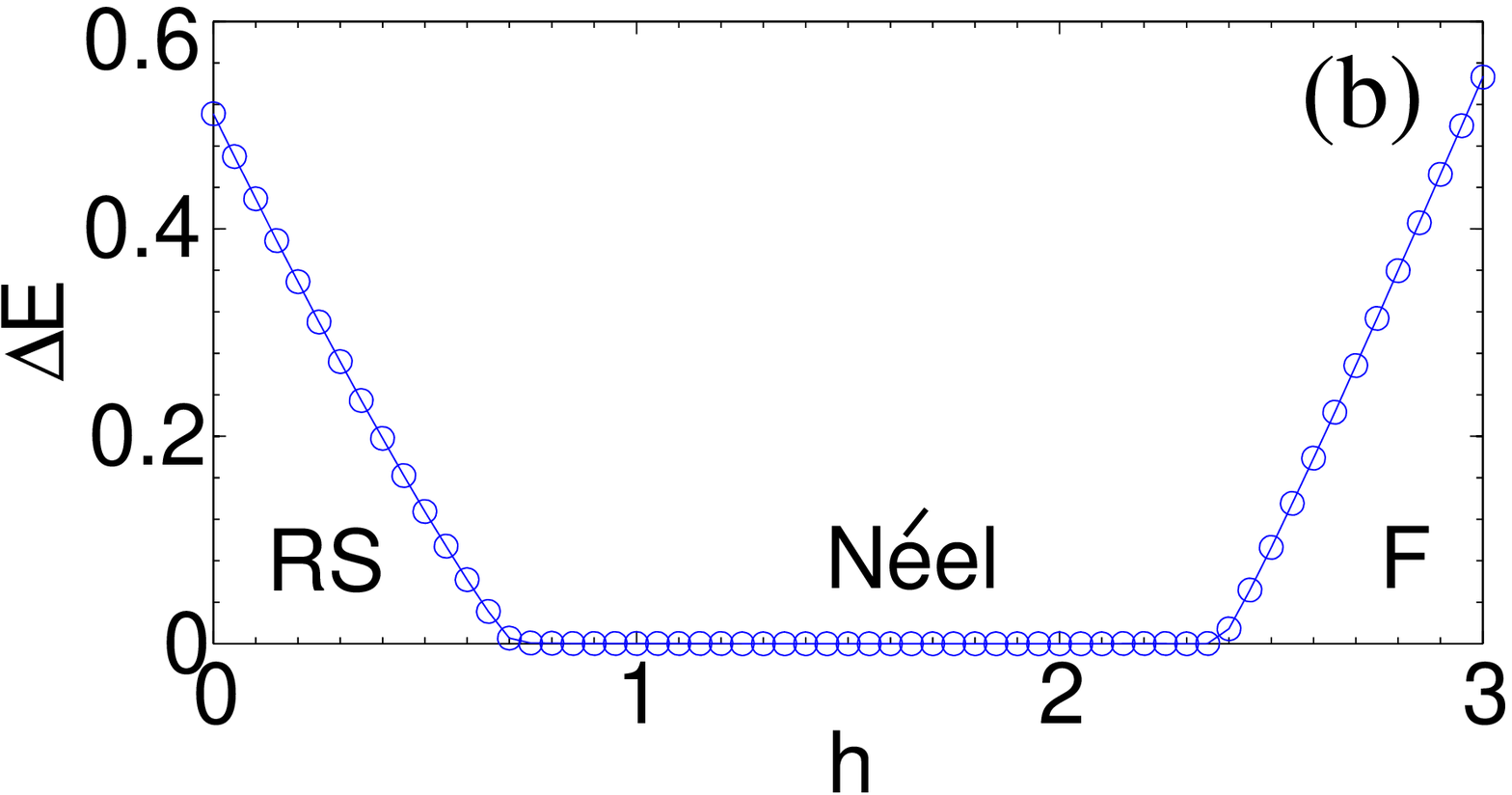}
\caption{ (a) Square N\'{e}el order as a function of the magnetic field along a cut through RS-N\'{e}el-F phases for $k_0^x=0$ and $k_0^y=\pm 3\pi/16$; 
(b) Behavior of the energy gap between the two lowest eigenstates as a function of the magnetic field across the RS-N\'{e}el-F phases. The N\'{e}el state is characterized 
by doubly degenerate ground states. The magnetic field and the gap are both measured in units of $J_{\parallel}=J_{\bot}$. 
The depicted results correspond to DMRG simulations with $L=48$ rungs. Finite size-effects near both phase transitions are very similar to the behavior depicted 
on Figs.~\ref{fig:squareNeel} (a) and (b) close to the N\'{e}el to F transition.}
\vspace*{-0.3cm}
\label{fig:squareNeelRung}
\end{figure}

We recall that for $k^y_0a=\pi/2$ a growing magnetic field does not introduce any phase transition 
but rather adiabatically connects RS and F phases. Since in the vicinity of $k_y^0a=0$ an intermediate N\' eel phase occurs, we
hence expect as a function of $k_0^ya$ and $h$ the presence of a N\' eel island inside an overall RS-F state.
Our numerical results confirm this expectation, as depicted in Fig.~\ref{fig:USOCrungs}.  
Since N\'{e}el order is spontaneous, in our numerical calculations we monitor 
\begin{equation}
n^2\equiv \lim _{|i-j|\gg 1}|(-1)^{i-j} \langle  S_{\alpha, i}^z   S_{\alpha,j}^z\rangle| .
\end{equation}
The magnetic field dependence of $n^2$ is illustrated in Fig.~\ref{fig:squareNeelRung}(a). 

We have studied as well the behavior of the excitation gap.  
The N\'{e}el state is characterized by a doubly-degenerate ground-state in the thermodynamic limit, 
whereas the RS and F states have unique gapped ground-states. 
Hence a simple way to obtain the boundary of the N\'{e}el state is to follow the closing of the gap 
between the ground-state and the first excited state (that becomes degenerate with the ground state in the thermodynamic limit in the N\'{e}el phase). 
We plot the behavior of the gap in Fig~\ref{fig:squareNeelRung}(b). 
The gaps close linearly with the magnetic field when approaching the quantum phase transition points, as expected from the Ising character.

 We note finally, that the vector product of two neighboring spins has a finite expectation value along rungs, $\langle [\mathbf{S}_{1,j} \times \mathbf{S}_{2,j} ]^z  \rangle \sim -\sin (2k^y_0 a) $, in all phases, becoming zero only deep in the F phase for large values of $h$.

\subsection{2-leg ladder with USOC along legs}
\label{subsec:USOC-legs}

\begin{figure}[ht]
\vspace*{0.3cm}
\includegraphics[width=8.0cm]{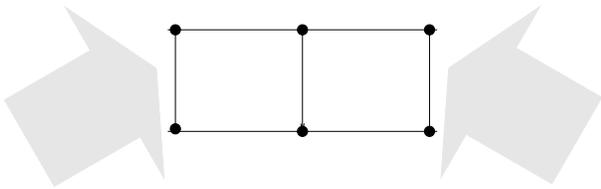}
\caption{ Raman lasers counter- propagating along the ladder legs result in an USOC as that discussed in subsection~\ref{subsec:USOC-legs}. 
The value of $k_x^0$ can be controlled by the angle between the laser propagation direction and the ladder legs.}
\vspace*{-0.3cm}
\label{fig:laserlegs}
\end{figure}

\begin{figure}[t]
\vspace*{0.3cm}
\includegraphics[width=0.8\columnwidth]{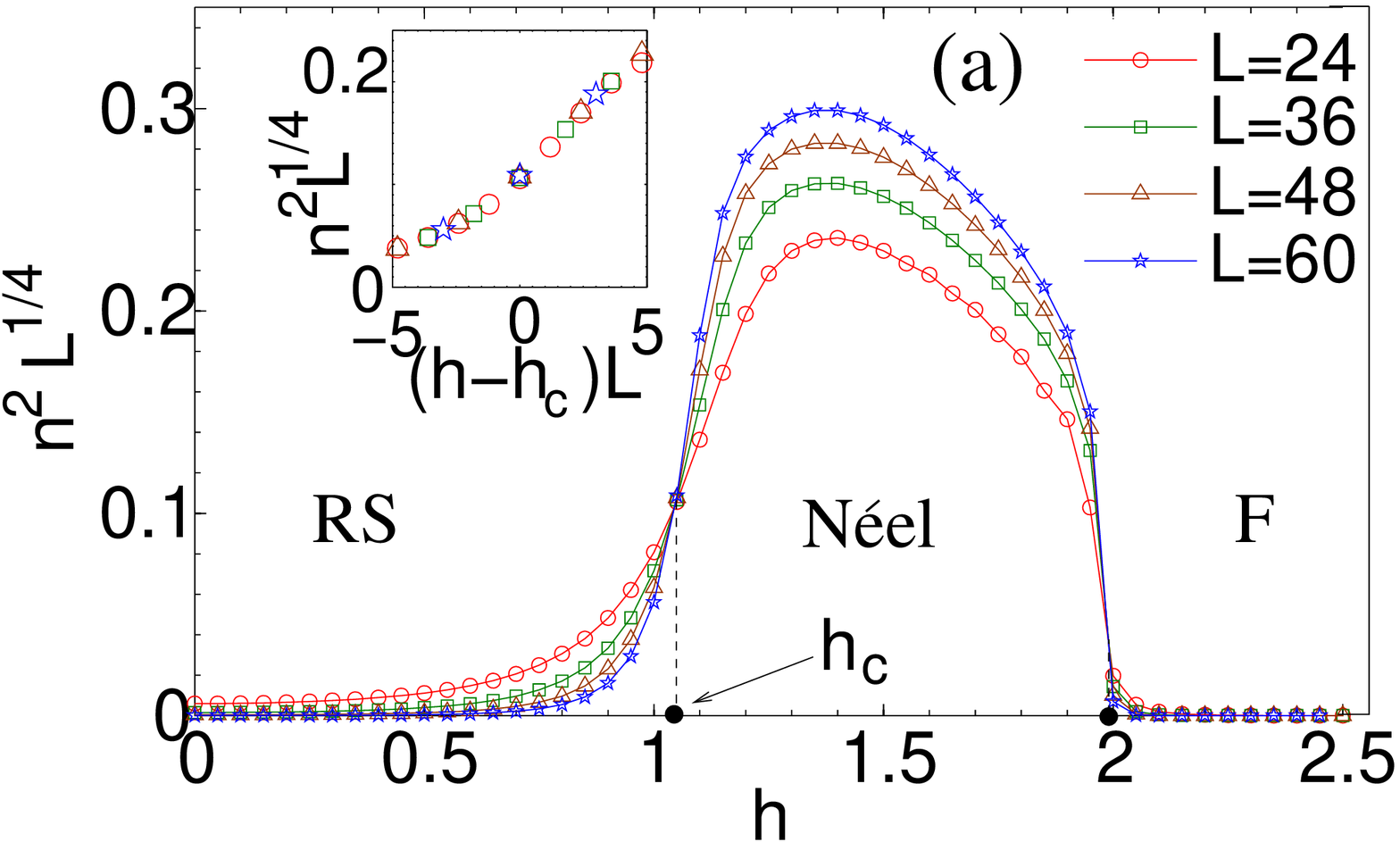}
\includegraphics[width=0.8\columnwidth]{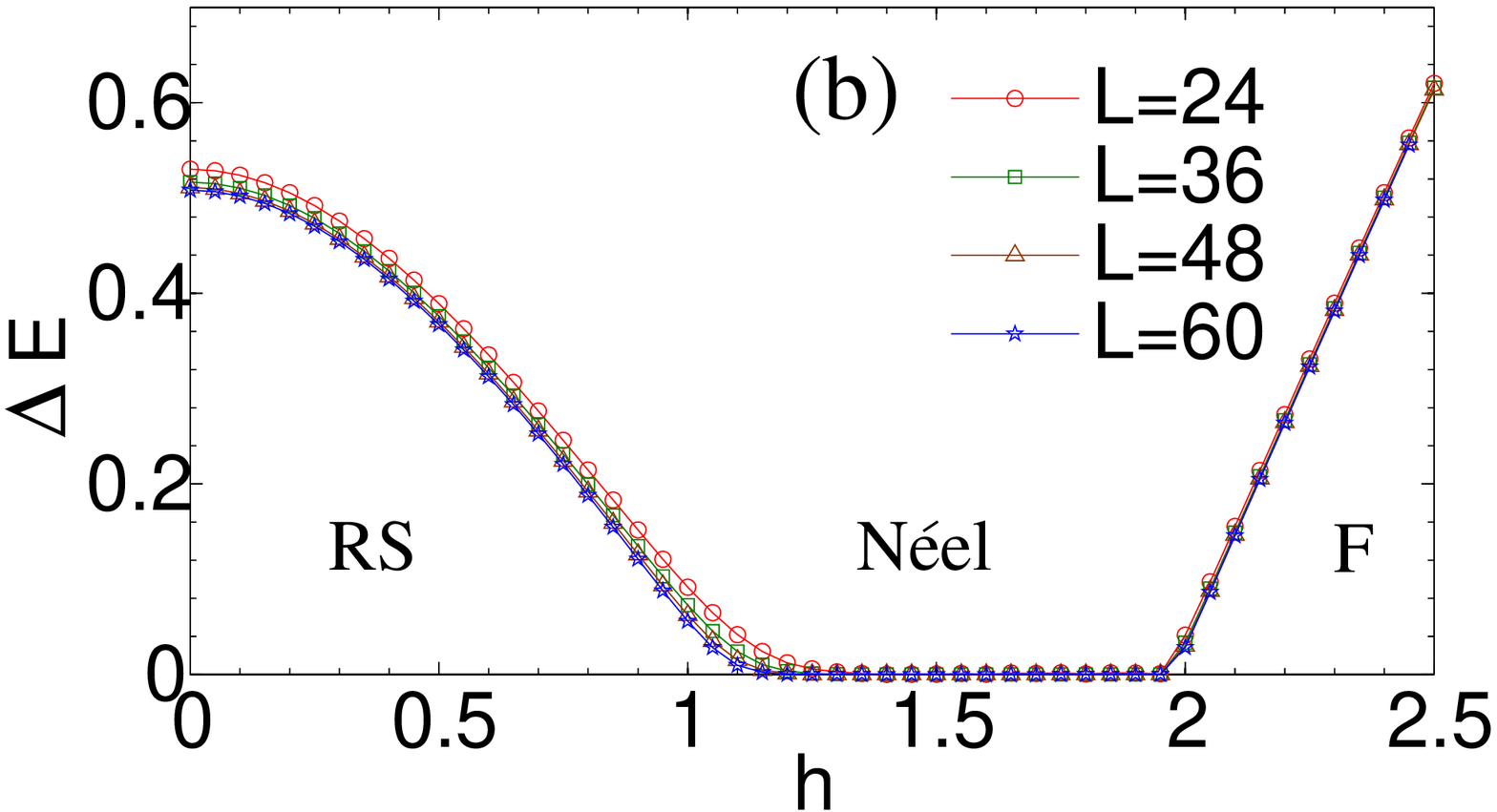}
\caption{ (Color online) (a) Square N\'{e}el order, $n^2$, as a function of the magnetic field along a cut through RS-N\'{e}el-F phases for $k_0^x=\pi/4$ and $k_y^0=0$ and for different system sizes. 
The inset shows the collapse of our numerical results for different system sizes on a single curve according to the Ising scaling. (b) Behavior of the energy gap between the two lowest eigenstates as a function of the magnetic field across the RS-N\'{e}el-F phases. The N\'{e}el state is characterized by doubly degenerate ground states. The magnetic field and the 
gap are measured in units of $J_{\parallel}=J_{\bot}$. The results displayed correspond to DMRG simulations for $L=24, 36, 48$ and $60$ rungs.  }
\vspace*{-0.3cm}
\label{fig:squareNeel}
\end{figure}

We consider now that the USOC is oriented along the ladder legs, as depicted in Fig.~\ref{fig:laserlegs}, and hence 
$k^y_0=0$ in  Eq.~(\ref{2D}). For $k^x_0=0$, the external magnetic field induces two consecutive C-IC phase transitions: 
a first one from RS into the gapless LL phase, and a second one from LL into the fully polarized F state. 
For $k_x^0\neq 0$ we may perform a gauge transformation similar to the ones discussed above, which for the maximal USOC, $k^x_0 a=\pi/2$, 
results in a model similar to Eq.~(\ref{RStransforme}) but in this case with a field that couples uniformly to spins belonging to the same rung and it alternates from rung to rung,
$-h \sum_{\alpha,j}\!(-1)^j \bar S_{\alpha,j}^x$.
Using bosonization in the weak rung-coupling limit, $J_{\bot}\ll J_{\parallel}$, and in opposite limit $J_{\bot}\gg J_{\parallel}$ employing strong rung-coupling expansion it has been determined that such magnetic field 
introduces Gaussian criticality between two gapped phases of the antiferromagnetic spin ladder~\cite{WangNersesyan}, 
that for our original spin variables corresponds to RS and F states. 

The difference at maximal USOC between the case with USOC along the ladder legs and that with USOC 
along the rungs can be easily understood in the limit $J_{\bot}\gg J_{\parallel}$. 
For the case of  USOC along the ladder legs the magnetic field couples uniformly to the spins on the same rung, and 
hence it favors a triplet state on each rung with both spins pointing in the same direction, which alternates from one rung to the next.
This state is orthogonal to the RS configuration. 
In contrast, in Eq.~(\ref{RStransforme}) the magnetic field couples in a staggered way to the spins in the same rung, and the ground-state favored by a strong magnetic field is not 
orthogonal to the RS state. As a result, for the USOC along rungs the RS state can be adiabatically connected to the F state, whereas 
for the USOC along legs this is not possible.

Based on the previous discussion we hence expect that the two C-IC phase transition points for $k_0^x a=0$ have to evolve into a 
single Gaussian point for $k_0^xa=\pi/2$. As for the case of USOC along rungs, we can employ bosonization to understand this evolution 
of the critical points  in the limit 
$J_{\bot}\ll J_{\parallel}$. The leading instability once the magnetic field suppresses the RS phase is again the EAA, 
however now in exchange interactions along the chains and DM anisotropy induces incommensurability~\cite{Citro}. 
In contrast to the relevant couplings produced by the USOC along the rungs, in 
Eq.~(\ref{effectivetwocomponent1}) the USOC along ladder legs produces a marginal perturbation, $\sim \cos {\sqrt{4\pi} \theta_-}\cos {\sqrt{4\pi} \theta_+}$. 
After mean-field decoupling between the symmetric and antisymmetric sectors the weak rung-coupling bosonic Hamiltonian for the USOC along 
legs is equivalent to Eq.~(\ref{effectivetwocomponent1}), and hence the ground-states and phase transitions will be similar to the previous case of 
USOC along rungs. Thus C-IC phase transition points evolve into Ising lines for $k_0^xa>0$ and at $k_x^0a=\pi/2$ these two Ising lines merge in a 
Gaussian criticality due to the enhanced symmetry.

\begin{figure}[t]
\vspace*{0.3cm}
\includegraphics[width=0.8\columnwidth]{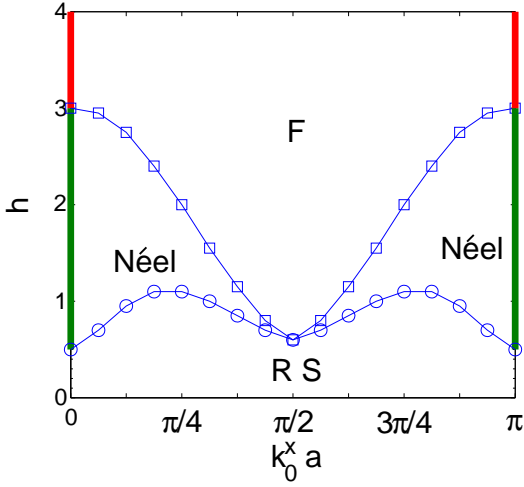}
\caption{(Color online) Numerical phase diagram for the USOC along the ladder legs, $k_0^y=0$. For $k_0^x=0 (\pi)$ there are two C-IC transitions with increasing magnetic field: 
a first one from RS to LL, and a second one from LL into the fully polarized state (both LL and fully polarized states are indicated by bold lines). For $0<k_0^x<\pi/2$ instead of the LL state a N\'{e}el state is realized, being separated from the RS and F states by Ising phase transition lines. These lines cross at $k_0^x=\pi/2$ resulting in a direct Gaussian 
transition from RS to F.  The magnetic field is in units of $J_{\parallel}= J_{\bot}$. Phase boundaries are obtained for the system with $L=48$ rungs.}
\vspace*{-0.3cm}
\label{fig:legRaman}
\end{figure}

Our numerical results for $n^2$ and the excitation gap are depicted in Figs.~\ref{fig:squareNeel}. 
Note that finite size effects are more pronounced at the RS to N\'{e}el transition, 
whereas for the N\'{e}el to F transition finite-size effects are negligible. 
For the RS to N\'{e}el transition we have carefully performed finite-size scaling of the order parameter and 
determined the critical field $h_c$ corresponding to the phase transition from the intersection of the order 
parameter curves for different system sizes. The collapse of order parameter for different system sizes 
in the vicinity of $h_c$ on the single curve according to the Ising law is depicted in the inset of Fig.~\ref{fig:squareNeel}(a).

Note finally that the vector product of two neighboring spins has finite expectation value along the chains
$\langle [\mathbf{S}_{\alpha,j} \times \mathbf{S}_{\alpha,j+1} ]^z  \rangle \sim -\sin (2k^x_0 a) $. 
Its magnetic field dependence is similar to the curve of Fig.~\ref{fig:chirality}(b), and it vanishes quickly in the F phase. 

The ground-state phase diagram for the USOC along the ladder legs is depicted in Fig.~\ref{fig:legRaman}. 
As mentioned above, the C-IC phase transition points (corresponding to $U(1)$ symmetry at $k^x_0=0$) transform into Ising 
transitions~(for  $0<k^x_0<\pi/2$ the system does not have continuous symmetry), and then they combine into a Gaussian point at  $k^x_0a=\pi/2$ (where $U(1)$ symmetry is revived).

So in both cases, USOC either along ladder rungs or along ladder legs, the system presents three possible phases, RS, N\' eel, and F. For a general orientation of the USOC and the ladder legs,
 $k_0^x \neq 0$ and $k_0^y\neq 0$, we hence expect these three phases as well.


\section{Two-leg ladder with non-abelian vector potential}
\label{sec:NonAbelian}

We consider at this point a non-Abelian vector potential of the form, ${\mathbf A}  =(-\hbar k_0^x \sigma^x, -\hbar k_0^y \sigma^y)$. 
Contrary to the case of USOC the magnetic field, $h$, is not necessary to ensure the non-trivial character of SOC. We hence consider 
the time-reversal symmetric case, $h=0$, and a balanced mixture of up and down spin fermions. 
The effective spin model in this case acquires the form:
\begin{eqnarray}
\label{2Dnonabelian}
&&H= J_{\parallel}\sum_{\alpha,j} \! \Big\{ \cos(2k^x_0 a) \mathbf{S}_{\alpha,j}  \mathbf{S}_{\alpha,j+1} \nonumber \\
&& +2\sin^2(k_0^x a)S^x_{\alpha,j} S^x_{\alpha, j+1}+\sin (2k^x_0 a)[\mathbf{S}_{\alpha,j} \times \mathbf{S}_{\alpha,j+1} ]^x \Big \} \nonumber\\
&& +J_{\bot}\sum_{j} \Big\{  \cos(2k^y_0 a) \mathbf{S}_{1,j}  \mathbf{S}_{2,j} \nonumber \\
&& +2\sin^2(k_0^y a) S^y_{1,j} S^y_{2,j} + \sin (2k^y_0 a)[\mathbf{S}_{1,j} \times \mathbf{S}_{2,j} ]^y \Big\}.  
\end{eqnarray}

\begin{figure}[ht]
\vspace*{0.3cm}
\includegraphics[width=0.9\columnwidth]{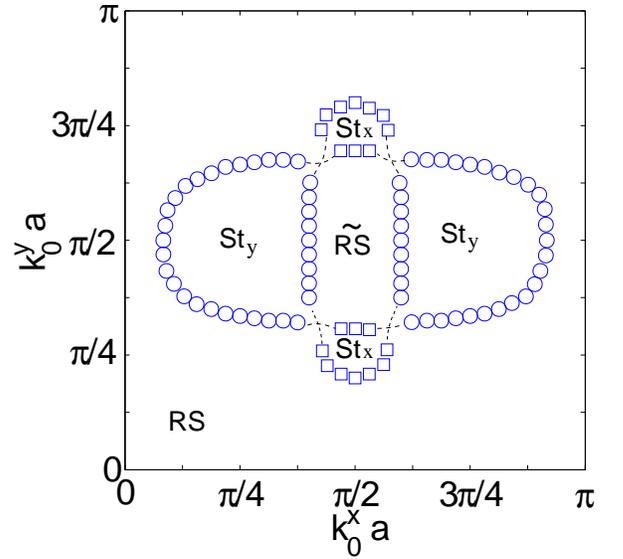}
\caption{  Ground states for spin-ladder with Rashba SOC, see text. The numerical results correspond to DMRG calculations for $L=48$ rungs. } 
\vspace*{-0.3cm}
\label{fig:2DPD}
\end{figure}
\begin{figure}[ht]
\vspace*{0.3cm}
\includegraphics[width=0.8\columnwidth]{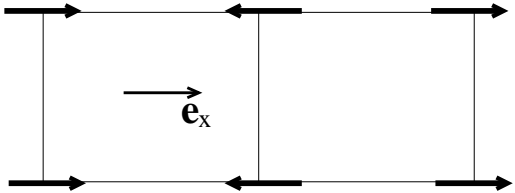}
\caption{  St$_y$ state configuration of a two-leg ladder with Rashba SOC. Spins are oriented along $\pm {\mathbf e}_x$ direction on odd rungs and along $\mp {\mathbf e}_x$ on even rungs. In  the St$_y$ state the vector products of nearest-neighbour spins behave as $\langle [\mathbf{S}_{\alpha,j} \times \mathbf{S}_{\alpha,j+1} ]^x  \rangle \sim - \sin (2k^x_0 a) $ and $\langle [\mathbf{S}_{1,j} \times \mathbf{S}_{2,j} ]^y  \rangle \sim 0 $. In the cartoon of St$_y$ state a nonzero value of $\langle [\mathbf{S}_{\alpha,j} \times \mathbf{S}_{\alpha,j+1} ]^x  \rangle$ quantity is not reflected.}
\vspace*{-0.3cm}
\label{fig:Neelferro}
\end{figure}
\begin{figure}[ht]
\vspace*{0.3cm}
\includegraphics[width=0.8\columnwidth]{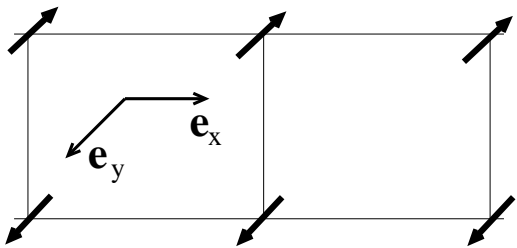}
\caption{  St$_x$ phase configuration of two-leg ladder with Rashba SOC. Spins are oriented along $\pm {\mathbf e}_y$ direction on one leg and along $\mp {\mathbf e}_y$ on another. In   St$_x$ phase the vector products of nearest-neighbour spins behave as $\langle [\mathbf{S}_{\alpha,j} \times \mathbf{S}_{\alpha,j+1} ]^x  \rangle \sim 0 $ and along rungs $\langle [\mathbf{S}_{1,j} \times \mathbf{S}_{2,j} ]^y  \rangle \sim -\sin (2k^y_0 a) $. }
\vspace*{-0.3cm}
\label{fig:Stripeferro}
\end{figure}

Our numerical results for the ground-state phase diagram are presented in Fig.~\ref{fig:2DPD}. 
In the vicinity of zero SOC, $ k_0^xa= k_0^ya=0$ the system is in the RS state. 
For the case of a maximal SOC, $ k_0^xa= k_0^ya=\pi/2$,  we may employ the canonical transformation
%
${\mathbf S_{\alpha,i}}\to  \mathbf{\tilde S}_{\alpha,i}   = U {\mathbf S_{\alpha,i}}  U^{\dagger}$, with 
\begin{equation}
U=  \prod_je^{-i\pi S^y_{1,j}} \prod_{\alpha,k=2j}e^{-i \pi S^x_{\alpha,k}},
\end{equation}
which transforms the Hamiltonian~(\ref{2Dnonabelian}) into an $SU(2)$ symmetric antiferromagnetic spin ladder Hamiltonian of the form of Eq.~(\ref{RStransforme}) with $h=0$.

Thus for  $ k_0^xa= k_0^ya=\pi/2$ the system is in the RS phase but in the gauge transformed spins~(and hence we denote it as $\tilde {\mathrm{RS}}$), 
with no long-range order and exponentially decaying correlation functions. In the strong rung coupling limit the ground state in gauge transformed variables is the rung-singlet product state Eq.~(\ref{rung-singlet}) with $\beta=1$, that for original 
variables transforms via $U$ to the direct product of $S^z=0$ components of the rung-triplets, also of the form of Eq.~(\ref{rung-singlet}) however with $\beta=-1$. Since the $\tilde {\mathrm{RS}}$ state is gapped it will occupy a finite region around $k_0^xa= k_0^ya=\pi/2$ point.

In the RS phase the vector product of two neighbouring spins has finite expectation value: along the chains
$\langle [\mathbf{S}_{\alpha,j} \times \mathbf{S}_{\alpha,j+1} ]^x  \rangle \sim - \sin (2k^x_0 a) $ and along rungs $\langle [\mathbf{S}_{1,j} \times \mathbf{S}_{2,j} ]^y  \rangle \sim -\sin (2k^y_0 a) $. In  $\tilde {\mathrm{RS}}$ phase the vector product of two neighboring spins is negligeably small.

Our numerical results reveal as well the appearance of striped phases with long range order where spins are ferromagnetically ordered in one direction and antiferromagnetically in the other. The case of ferromagnetic order along the rung~(St$_y$ phase) is best understood in the vicinity of $(k_0^xa,k_0^ya)=(\pi/4,\pi/2)$ point, where $\cos(2k_0^x a) =0$, and  $\cos(2k_0^y a) =-1$. For these parameters the coupling along the rung  $S_{1,j}^xS_{2,j}^x$ is ferromagnetic, whereas intra-leg coupling  $S_{\alpha,j}^xS_{\alpha,j+1}^x$ is antiferromagnetic, which results in the St$_y$ configuration observed in our numerical calculations~(Fig.~\ref{fig:Neelferro}). 

We may understand in a similar way the appearance of the St$_x$ phase, see Fig.~\ref{fig:Stripeferro}, analyzing the behavior in 
 the vicinity of $(k_0^xa,k_0^ya)=(\pi/2,\pi/4)$, which is characterized by a ferromagnetic $S_{\alpha,j}^yS_{\alpha,j+1}^y$ coupling along legs, and an antiferromagnetic 
 $S_{1,j}^yS_{2,j}^y$ exchange along rungs. St$_x$ and St$_y$ states are dual to each other with respect to the interchange of leg and rung directions and $S^x$ and $S^y$ components.
Our numerical simulations suggest that similarly to the USOC case all phase transitions for the case of the non-Abelian vector potential are 
of second-order Ising nature. This is natural, since the system does not enjoy in general any continuous symmetry, 
and striped phases break spontaneously discrete $Z^2$ symmetries: St$_y$ breaks translation symmetry along the chains 
direction whereas St$_x$ breaks the parity symmetry associated with the exchange of ladder legs. Both striped phases break as well time reversal symmetry.

The RS and $\tilde {\mathrm{RS}}$ phases present different parity symmetry for an odd number of rungs, whereas the RS phase is antisymmetric the $\tilde {\mathrm{RS}}$ is symmetric; As a result both phases cannot connect adiabatically. We could not determine numerically whether RS and $\tilde {\mathrm{RS}}$ states can be connected adiabatically for an even number of rungs in the parameter space $(k_0^x,k_0^y)$ in Fig.~\ref{fig:2DPD}. In particular, the string order, defined for the pair of spins across the ladder diagonal, is finite for both RS and $\tilde {\mathrm{RS}}$ states and vanishes in the striped phases. However in the thermodynamic limit we expect the behavior of odd and even number of rungs to converge, and hence it is most likely that in the model given in Eq.~(\ref{2Dnonabelian}) RS and  $\tilde {\mathrm{RS}}$ states are always separated by a phase transition~(indicated by dashed lines in Fig.~\ref{fig:2DPD}).

\section{Conclusions}
\label{sec:Conclusions}

In our work we have discussed the quantum spin phases and the associated quantum phase transitions for a two-component Fermi lattice gas, focusing on the 
case of a two-leg ladder-like lattice at half-filling, a minimal system to study the non-Abelian character of the vector potential. We have shown that for an USOC along the ladder rungs an N\'{e}el state phase is located within a RS-F phase, in which a rung-singlet may be adiabatically connected to a ferromagnetic phase in the parameter space of $h$ and the SOC. In contrast, for the USOC along the ladder legs the RS and F states cannot be adiabatically connected, and are separated by an intermediate  N\'{e}el state, 
which disappears at a maximal SOC to lead to a direct Gaussian RS-F quantum phase transition. The case of a Rashba-like SOC is characterized by the appearance of rung-singlet and striped phases. Compared to the classical spin phases predicted for fermions on a square lattice with SOC~\cite{Radic} 
only the striped configurations of the 2D lattice have identical quantum counterparts on the ladder. On the contrary, the N\'{e}el and spiral waves are substituted by gapped rung-singlet states,  whereas non-coplanar configurations such as  vortex/antivortex textures are not stabilized.

\acknowledgements

We thank A. K. Kolezhuk and S. Manmama for discussions. This work has been supported by QUEST (Center for Quantum Engineering and Space-Time Research) and DFG Research Training Group (Graduiertenkolleg) 1729.

\end{document}